\documentclass[apj,revtex4]{emulateapj}
\usepackage{url}
\usepackage{natbib}
\usepackage{amssymb}
\usepackage{amsmath,bm}
\usepackage{pstricks}
\usepackage{color}
\usepackage{graphicx}
\usepackage{slashed}
\usepackage{relsize}
\usepackage{multirow}
\usepackage{hyperref}
\shorttitle{Gravitational microlensing by neutron stars and radio pulsars}
\shortauthors{Dai, Smith, Lin, Yue, Hobbs \& Xu}
\begin{document}

\title{Gravitational Microlensing by Neutron Stars and Radio Pulsars: Event Rates, Timescale Distributions, and Mass Measurements}
%\title{Properties of Microlensing Events Caused by Neutron Stars and Mass Measurements}

\author{S. Dai$^{1,2}$, M. C. Smith$^3$, M. X. Lin$^1$, Y. L. Yue$^4$, G. Hobbs$^2$, R. X. Xu$^{1,5}$}
\affil{$^1$School of Physics and State Key Laboratory of Nuclear Physics and Technology, Peking University, Beijing 100871, China}
\affil{$^2$CSIRO Astronomy and Space Science, Australia Telescope National Facility, Box 76 Epping NSW 1710, Australia}
\affil{$^3$Shanghai Astronomical Observatory, Chinese Academy of Sciences, Shanghai 200030, China}
\affil{$^4$National Astronomical Observatories, Chinese Academy of Sciences, Beijing 100012, China}
\affil{$^5$Kavli Institute for Astronomy and Astrophysics, Peking University, Beijing 100871, China}
\email{daishi@pku.edu.cn}
\email{msmith@shao.ac.cn}

\begin{abstract}
We investigate properties of Galactic microlensing events in which a
stellar object is lensed by a neutron star.
For an all-sky photometric microlensing survey, we determine the
number of lensing events caused by $\sim10^{5}$
potentially-observable radio pulsars to be $\sim0.2\ \rm{yr^{-1}}$ for
$10^{10}$ background stellar sources. We expect a few detectable
events per year for the same number of background sources from an
astrometric microlensing survey.
We show that such a study could lead to precise measurements of radio
pulsar masses. For instance, if a pulsar distance could be constrained
through radio observations, then its mass would be determined with a
precision of $\sim10\%$.
We also investigate the time-scale distributions for neutron star
events, finding that they are much shorter than had been previously
thought.
For photometric events towards the Galactic centre that last
$\sim15$\,days, around 7\% will have a neutron star lens. This
fraction drops rapidly for longer time-scales. Away from the bulge
region we find that neutron stars will contribute $\sim40\%$ of the
events that last less than $\sim10$\,days. These results are in
contrast to earlier work which found that the maximum fraction of 
neutron star events would occur on time-scales of hundreds of days. 
\end{abstract}


\section{Introduction}

Neutron stars (NSs), as the densest observable objects in the Universe,
allow us to study diverse phenomena in
astrophysics and also probe fundamental physics at extremely high 
density. For instance, determining whether pulsar-like compact objects
are NSs or quark stars would have profound implications on the physics of
condensed matter and on the nature of the strong interaction~\citep[e.g.,][for a review]{Weber,Xu10}.
% and the QCD phase transition

Determining NS masses is essential for constraining properties 
of their inner structure.
The existence of high-mass NSs places strong constraints on the 
equation of state, while extremely low-mass NSs help distinguish 
self-bound quark stars from gravity-bound NSs.
Recent precise high mass measurements of PSRs J0348$+$0432 and J1614$-$2230
significantly constrained the equation of state of NSs and improved our 
understanding of pulsar-like compact objects~\citep{Anton,Demorest,Ozel2010,Lai2011}.
However, up to now, all precise mass measurements are from observations
of binary pulsars. Determining masses for isolated NSs is challenging, 
but knowing the mass distribution for such NSs is important as it may be 
different from binary NSs.

Gravitational microlensing has been suggested as a 
way to measure the masses of isolated NSs~\citep[e.g.,][]{Dai,Schwarz02,Horvath96}.
There are two kinds of microlensing phenomena that occur when the 
line-of-sight to a distant object passes close to an intervening massive object.
The first is known as photometric microlensing, which is observed as the temporal 
brightening of the background star. For photometric microlensing events the photometric 
magnification, $A$, depends on the angular separation between the lens and the 
source in units of the angular Einstein radius, $u=\theta_{\rm{sl}}/\theta_{\rm{E}}$, 
as $A\sim 1+2/u^{4}$~\citep{mao}. A typical angular Einstein radius
for a stellar microlensing event in the Galaxy is approximately 1\,mas. 
The second is known as astrometric microlensing and is observed as the angular shift, 
$S$, of the apparent position of the background star. For angular separations larger 
than the angular Einstein radius, $\theta_{\rm{E}}$, the shift $S \sim \theta_{\rm{E}}/u$~\citep[e.g.,][hereafter BE02]{Bel}.  
Therefore, the cross-section for astrometric events will be much larger than for 
photometric events because the detectability of astrometric events is not as 
strongly dependent on $u$. However, it is currently impractical to carry out 
large area astrometric microlensing surveys.

The Optical Gravitational Lensing Experiment (OGLE)~\citep{Udalski} is an ongoing 
photometric microlensing survey.    
The third phase of this project, OGLE-III, monitored around $3\times10^{8}$ 
stars in the Galactic bulge~\citep{Szymanski}, as well as fields in
the disc and towards the Magellanic Clouds. The current phase of
the OGLE project, OGLE-IV, is covering a significantly larger area and
detecting many more events than OGLE-III.
In the near future the Large Synoptic Survey Telescope
(LSST)~\citep[for more detailed discussion see the LSST Science
Book;][]{lsst} will also detect large numbers of microlensing events.

There are currently no wide-area surveys for astrometric microlensing
events and none planned in the foreseeable future. However, if an
astrometric microlensing event could be
predicted then a specific observing campaign could monitor the particular area of sky in 
which the event will occur. It is possible to determine astrometric parameters independently 
for NSs that can be observed as radio pulsars.  For such pulsars it is therefore possible 
to predict times of microlensing events. A further advantage is that the distance to 
the pulsar can be independently estimated. The properties of such a microlensing event 
would therefore be well constrained and would provide a direct measurement of the mass of
the pulsar lens. Currently only a relatively small number of radio pulsars are known.  
However, future radio telescopes, such as the Five-hundred-meter Aperture Spherical 
Telescope~\citep[FAST,][]{Nan} and the Square Kilometre Array~\citep[SKA,][]{Johnston2007}, 
will discover a significant fraction of all the potentially-observable radio pulsars in our galaxy and will
be able to determine the astrometric parameters for many of these pulsars precisely.

In the first part of this paper we study photometric NS lensing events. We calculate 
the expected event rates, the characteristic time-scale of NS events and determine the 
fraction of total events caused by NSs.  Previous studies~\citep[e.g.,][hereafter WM05]{Sartore10,wood}  
were not based on the most up-to-date spatial and velocity distributions for the 
NS population~\citep[e.g.,][]{hobbs,Lorimer06,Faucher}. NS lensing events are only 
a small fraction of the total number of events detectable in a lensing survey. We 
show that knowledge of the expected characteristic time-scale of NS events can significantly 
improve identification of such events.  

In the second part of the paper we make predictions for astrometric microlensing 
events caused by NSs. We provide expected events rates and determine the likely precision 
that could be achieved for NS masses. This has not been investigated previously (previous 
studies concentrated on stellar lenses~e.g., BE02). We conclude by discussing 
the implications of our results and providing predictions for future surveys.

\section{Properties of photometric microlensing caused by Neutron Stars}

In this section, we investigate photometric microlensing events caused by NSs. 
NSs can lens emission from Galactic stellar objects and also from background galaxies.  
As \citet{Tian} have already studied NSs lensing faint background galaxies, we 
concentrate on Galactic background stellar objects. Our work follows the 
procedure described in WM05. 
In Section $2.1$, we describe the spatial and velocity 
distribution of the NS population and the stellar models used in 
our simulations.
We then present the event rates, time-scale distributions of microlensing
events caused by NSs and their fractional contribution to the Galactic event
rates.

\subsection{NS and stellar population distributions}

\subsubsection{The distribution of the NS population}

We assume that the NS number density is proportional to that of radio pulsars.
In cylindrical coordinates originating at the Galactic Center (GC), the NS 
number density is expressed as
\begin{equation}
\begin{split}
\rho(R,z) & =A\left(\frac{R+R_{1}}{R_{\odot}+R_{1}}\right)^{\rm{a}}\exp\left[-b\left(\frac{R-R_{\odot}}{R_{\odot}+R_{1}}\right)\right]\\
          & \times\exp\left(-\frac{|z|}{E}\right),
\end{split}
\end{equation}
where $R_{1}=0.55$ kpc, $a=1.64$, $b=4.01$, $E=330$ pc and the distance from the
Sun to the GC is set to be $R_{\odot}=8.0$ kpc.
This assumes the radial distribution of NSs as suggested by~\citet{Yusifov}. Such a 
radial distribution was also used in~\citet{Faucher}, and allows us to use their 
prediction for the number of potentially-observable pulsars (i.e. NSs that are 
radio-loud and beamed toward us). For the $z$ distribution we use the scale height 
obtained by~\citet{Lorimer06}.
The total number of NSs in the Galaxy is still unclear. Following recent 
estimations~\citep{Keane}, we normalize the parameter $A$ to 
give a total NS number of $10^9$. For potentially-observable pulsars  
$A$ is normalized to give a total number of observable pulsars of $120,000$~\citep{Faucher}. 
We note that this number does not account for the population of Rotating Radio 
Transients~\citep[RRATs, ][]{McLaughlin06}. Their inclusion would significantly 
increase the number of observable NSs~\citep{Keane}.

We adopt a Gaussian model with $\sigma=290\ \rm{km\ s^{-1}}$ \citep{Faucher} for 
the distribution of the components of the NS velocities.
As will be defined in Section 2.2, the time-scale and event rate 
is inversely proportional and proportional to the lens-source relative transverse 
velocity respectively, therefore increasing or decreasing the $\sigma$ of the NS 
velocity distribution will simply rescale the time-scale and event rate. 
Although an exponential model is preferred in~\citet{Faucher}, the Gaussian
model has also been shown to well match the velocity distribution of radio 
pulsars~\citep{hobbs} and provides a straightforward distribution for us to use 
in our calculations.
In the analysis below we use the Gaussian model by default, but for comparision 
we also give some results based on the exponential model.

\subsubsection{Stellar models}

In order to determine the properties of the background stars we consider 
stellar populations in the bulge and in the disk.
For the bulge distribution of stellar objects, we use the E2 model in~\citet{Rattenbury}.
We truncate the bulge at a corotation radius $R_{C}=3.5\ \rm{kpc}$~\citep{Bissantz}. 
The model is normalized to the observed value of the optical depth towards the 
Central Galactic Region~\citep{Calchi,popowski}.
The disk distribution of stellar objects is relatively well constrained, and we use 
the local vertical density model of~\citet{zheng}, as extended to the whole 
Galactic disk by \citet{han}.

We also need models of the lensing objects (both NSs and other objects). 
For the lens mass function, we follow an identical procedure to that 
described in WM05. This is based on a two part power-law distribution,
\begin{equation}
\frac{\rm{d}N}{\rm{d}M}=k\left(\frac{M}{M_{\rm{brk}}}\right)^{\rm{\alpha}},
\end{equation}
where $M_{\rm{brk}}=0.7\ \rm{M_{\odot}}$, $\alpha=-2.0$ for $M>M_{\rm{brk}}$,
and $\alpha=-1.3$ for $M\leq M_{\rm{brk}}$. 

We adopt the kinematic model of WM05, which describes the 
velocities of the lenses, sources and the observer. The lens-source 
relative transverse velocity, $\upsilon$, is calculated according 
to Eq. 7 and 8 in WM05. The conversion of velocities from the 
Galactocentric cylindrical coordinates into solar-centric spherical
coordinates is given by Eq. 15 in~\citet{wang}.

\subsection{The photometric microlensing model}

The time-scale of a microlensing event is the
time taken for a source to cross the Einstein radius of the lens,
$r_{\rm{E}}$, and is defined as~\citep{Paczynski1996}
\begin{equation}
\label{te}
t_{\rm{E}}=\frac{r_{\rm{E}}}{\upsilon},\
r_{\rm{E}}=\sqrt{\frac{4GM}{c^{2}}\frac{D_{\rm{d}}(D_{\rm{s}}-D_{\rm{d}})}{D_{\rm{s}}}},
\end{equation}
where $G$ is the gravitational constant, $c$ is the speed of light, $D_{\rm{s}}$ is 
the source distance and $D_{\rm{d}}$ is the lens distance.

The event rate, $\Gamma$, is defined as the number of photometric microlensing 
events per unit time for a given number of background sources.
The event rate depends upon the background source number density, $\rho_{\rm{s}}(D_{\rm{s}})$, 
and the mass density of lensing objects, $\rho_{\rm{m}}(D_{\rm{d}})$.
Following~\citet{kiraga} and WM05, we assume that the fraction of 
stars brighter than some luminosity $L$ is proportional to $L^{\beta}$, and 
then the event rate is expressed as:
\begin{equation}
\label{rate}
\begin{split}
\Gamma=&\frac{4G^{1/2}}{c}\int_{0}^{\infty}\rm{d}D_{\rm{s}}D_{\rm{s}}^{2+2\beta}\rho_{\rm{s}}(D_{\rm{s}})\\
	     &\frac{\int_{0}^{D_{\rm{s}}}\rm{d}D_{\rm{d}}\rho_{\rm{m}}(D_{\rm{d}})\upsilon[D_{\rm{d}}(D_{\rm{s}}-D_{\rm{d}})/MD_{\rm{s}}]^{1/2}}{\int_{0}^{\infty}\rm{d}D_{\rm{s}}D_{\rm{s}}^{2+2\beta}\rho_{\rm{s}}(D_{\rm{s}})},
\end{split}
\end{equation}
where $\upsilon$ is the lens-source relative transverse velocity.

We use $\beta$ to model numerous phenomena that are not explicitly included in the 
equation. To account for the luminosity function of the background sources, the optical 
survey sensitivity and the Galactic extinction, we set $-3\leq\beta\leq-1$~\citep{kiraga}. 
We also calculate our results with $\beta=0$, which corresponds to the maximum possible 
event rate. This is a simplified approach, but has been applied by previous studies on 
the Galactic microlensing event rate and is valid as we only report basic properties of 
NS lensing events and do not attempt to predict the exact number of events detected with 
a particular set of survey parameters. 
Although extinction will clearly affect our estimates for the event rates, implementing 
this in a consistent manner is a very complex task~\citep[see, for example,][]{kerins}. 
However, since extinction results in a change in the slope of the luminosity function, 
this can be reflected in the choice of the $\beta$ parameter, i.e., large negative values of 
beta can be used to account for high extinction.
We consider stellar objects in the Galactic disk and in the bulge as background sources. 
As the disk stars can have small distances, setting $-3\leq\beta\leq-1$ will result in 
unreasonably small event rates. We account for this by always setting $\beta = 0$ for 
disk sources within $4.5$\,kpc from the observer.

To calculate the event rates and mean time-scales, we use the adaptive Monte Carlo 
integration algorithms incorporated in the GNU Scientific Library\footnote{\url{http://www.gnu.org/software/gsl/}}.
In order to improve the calculation precision we use a 
sampling number of $10^{7}$ over the multi-dimensional volume when we 
carry out the Monte Carlo integrations.

\subsection{Photometric microlensing results}

\subsubsection{Event rates}

Our results give a maximum possible all-sky averaged event rate
($\beta = 0$) of $4.2\times10^{-7}\ \rm{yr^{-1}}$ for NS
lenses. Towards the GC and the Baade's window
(BW)\footnote{$(l,b)=(1.16^{\circ},-2.75^{\circ})$, $l$ and $b$ are
the Galactic longitude and latitude.} region this maximum event rate
is $6.7\times10^{-7}\ \rm{yr^{-1}}$ and $3.6\times10^{-7}\ \rm{yr^{-1}}$,  
respectively. Setting $\beta=-1,-2,-3$, the all-sky averaged event
rate is $1.3\times10^{-7}\ \rm{yr^{-1}}$,
$7.5\times10^{-8}\ \rm{yr^{-1}}$ and $7.4\times10^{-8}\ \rm{yr^{-1}}$,
respectively. In Table \ref{eventRate} we present the all-sky averaged
event rate, rates for $\mid b\mid<5^{\circ}$ and $\mid
b\mid>5^{\circ}$ and rates towards the GC and BW, for
$\beta=0,-1,-2,-3$ for 120,000 potentially-observable radio
pulsars. Since we have assumed that the population model for radio
pulsars is identical to that of NSs, except for the normalization, this
means that the event rates for potentially-observable radio pulsars is
a factor of $10^9/\left(1.2\times10^5\right)$ smaller than that of NSs.
%The event rates reduce significantly as $\beta$ decreases because fewer distant sources contribute to the rate.
Using the exponential	model proposed by~\citet{Faucher}, we obtained an
all-sky averaged event rate for potentially-observable radio
pulsars of $4.2\times10^{-11}\ \rm{yr^{-1}}$ ($\beta=0$), compared 
with $5.0\times10^{-11}\ \rm{yr^{-1}}$ for the Gaussian model.

\begin{figure}
\begin{center}
  \includegraphics[width=3.5 in]{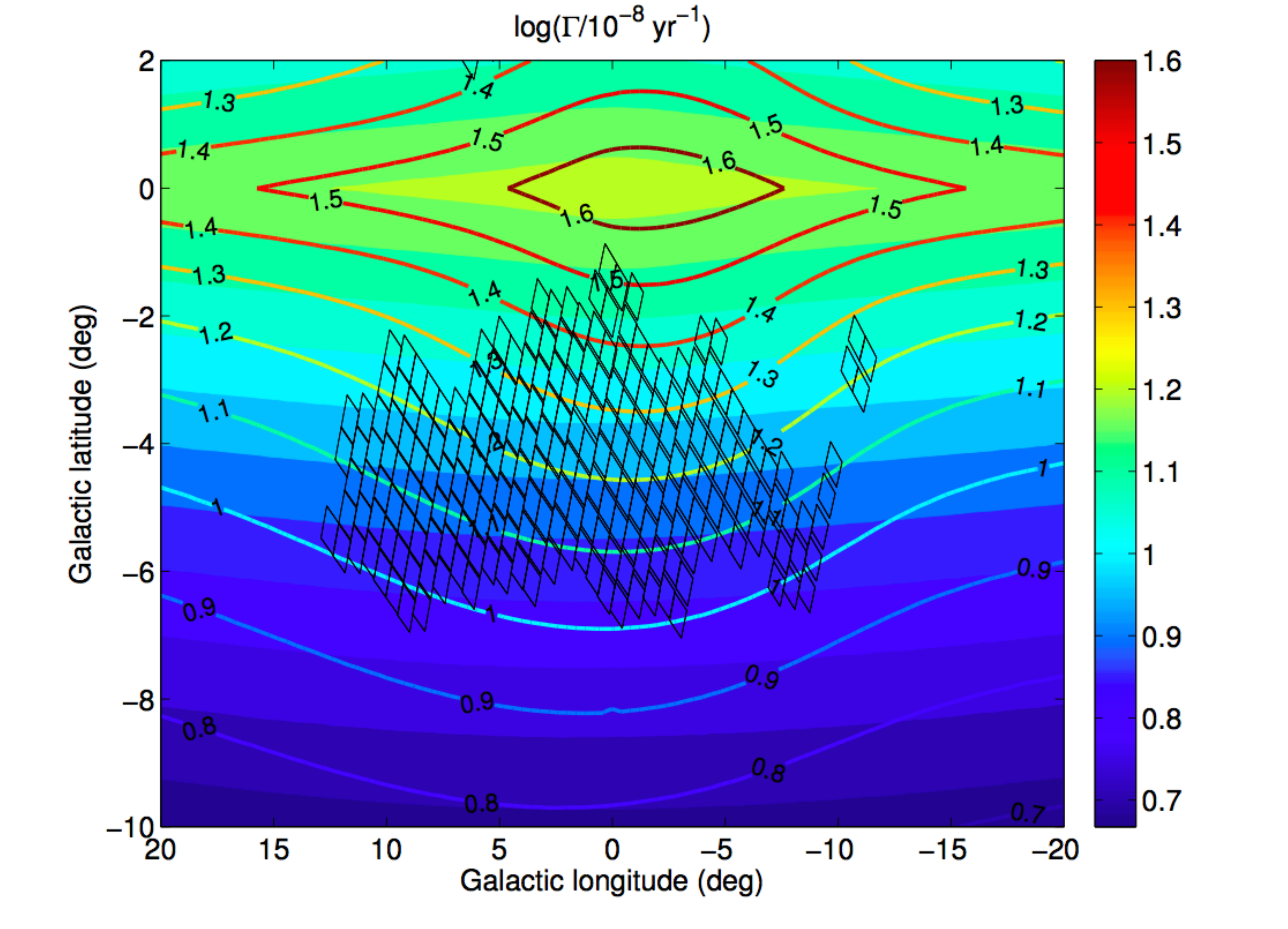}
\caption{Map of the microlensing event rates caused by NSs, towards the central Galactic
region. Contours with solid lines represent results for $\beta=-1$. Contours filled with
colors represent results for $\beta=-2$. OGLE-III fields are shown with diamond symbols. Contour levels
represent the event rate in logarithmic scaling, $\log(\Gamma/10^{-8}\ \rm{yr^{-1}})$. }
\label{map_event}
\end{center}
\end{figure}

We show in Fig.~\ref{map_event}, the expected NS event rates in the
central Galactic region. The contour levels represent the event rate
using a logarithmic scaling, $\log(\Gamma/10^{-8}\ \rm{yr^{-1}})$.
Two sets of contours are shown. Contours for $\beta = -1$ are shown with solid lines.  
Contours for $\beta = -2$ are shown as shaded regions. As expected, the event rate is higher 
on the Galactic plane and decreases at higher Galactic latitudes. As $\beta$ varies from $-1$ 
to $-2$, the distribution of the event rates becomes flatter, and the rate decreases much more 
significantly on the Galactic plane compared to that at a higher Galactic latitude region. 

The reduction in event rates as $\beta$ decreases and the flattening of the event rate distribution 
can be explained as:
\begin{itemize}
\item Decreasing $\beta$ implies that fewer distant sources contribute to the event rate. Therefore 
the total event rate will decrease as $\beta$ decreases.
\item Towards the GC the stellar density is higher than in the Galactic disk. 
Decreasing $\beta$ therefore has a larger effect in the GC region than elsewhere.
\item Sources in the bulge have a high velocity dispersion. Therefore decreasing $\beta$ effectively 
removes more high-velocity stellar objects in the bulge compared with the disk.
\item Sources on the far side of the Galactic disk rotate in the opposite direction to the observer.  
These sources therefore contribute a significant number of events as their relative velocities to the 
NS lenses are high. Decreasing $\beta$ removes these distant background sources.
\end{itemize}

%This flattening can be explained by the larger number density and the higher velocity dispersion for stellar objects in the Galactic bulge.

OGLE-III fields\footnote{data from \url{http://ogle.astrouw.edu.pl/}}
are shown in Fig.~\ref{map_event} with diamond symbols.
For $340\times10^6$ background sources~\citep{Szymanski} and taking
$\beta = -1$, we predict there should have been around 50 NS events
per year and about a 6\% chance that there has been an event from a
potentially-observable radio pulsar during the 9-year duration of the
survey.
The current phase of the OGLE project (OGLE-IV) started in
2010 and this is now detecting around four times as many microlensing
events per year in the bulge compared to
OGLE-III\footnote{\url{http://ogle.astrouw.edu.pl/ogle4/ews/ews.html}}.
This could therefore detect around two hundred NS events per
year and, given the event rate for potentially-observable radio
pulsars, there is a $\sim23$ per cent chance it will detect one by the
end of the decade.
For future deeper and wider optical surveys the prospects for discovering pulsar microlensing 
events are promising. For example, LSST is expected to monitor $\sim10^{10}$ background Milky Way 
sources~\citep{ivez12a}. Taking $\beta = -1$, we predict there would be $\sim1000$ events caused 
by NSs per year and $\sim1.5$ events caused by potentially-observable pulsars per decade.

Note that our photometric event rates are likely to underestimate the
actual number of observable events. This is because our calculations
are based on the definition of a microlensing event as one for which
the lens passes within one Einstein radius of the source,
corresponding to an amplification of 0.32\,mag. However, in practice
it is possible to detect events with much weaker amplification. If we
instead define an event as one for which the amplification is
0.1\,mag, then the event rate will be $\sim70$ per cent higher. There
is also the fact that the actual number of source stars may be
underestimated due to blending from unresolved source stars
\citep[e.g.][]{Smith07}.

%
%%%%%%%%%%%%%%%%%%%%%%%%%%%%%%%%%%%%%%%%%
%%%%%%%%%%%%%%%%%%%%%%%%%%%%%%%%%%%%%%%%%%
%

\begin{table}
\begin{center}
\caption{The event rates caused by potentially-observable radio
pulsars, in units of $10^{-11}\ \rm{yr^{-1}}$. To obtain the event
rate for NS lenses, the numbers in this table should be multiplied by
a factor of $10^9/\left(1.2\times10^5\right)$.}
\label{eventRate}
\begin{tabular}{lccccc}
\hline
    $\beta$    &      All-sky  &   $\mid b\mid<5^{\circ}$  & $\mid b\mid>5^{\circ}$  & GC  &  BW        \\
%               &    \multicolumn{5}{c}{($\times10^{-11}\ \rm{yr^{-1}}$)}                                \\
\hline
      0        &      5.0      &    5.7    &  1.3  &  8.0 & 4.4   \\
      -1       &      1.5      &    2.0    &  0.5  &  5.5 & 2.8   \\
      -2       &      0.9      &    1.2    &  0.4  &  2.0 & 1.4   \\
      -3       &      0.9      &    1.2    &  0.4  &  1.9 & 1.3   \\
\hline
\end{tabular}
\end{center}
\end{table}

\subsubsection{Time-scale distributions}

%%%%%%%%%%%%%%%%%%%%%%%%%%%%%%%%%%%%%%%%%
\begin{figure}
\begin{center}
  \includegraphics[width=3.5 in]{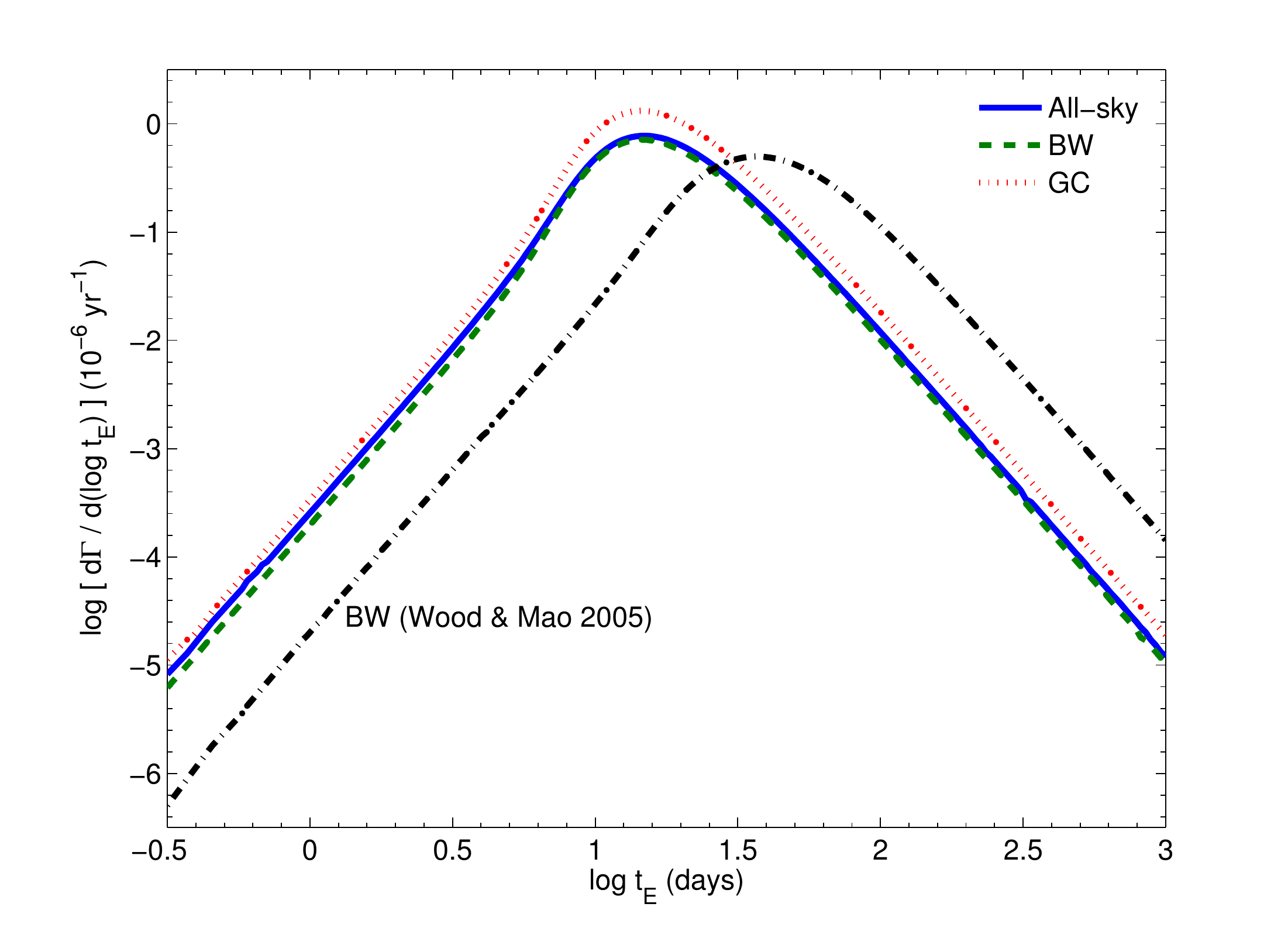}
\caption{Microlensing event rates caused by NSs as a function of time-scale.
The solid, dotted and dashed lines represent the all-sky averaged event rate and the
event rates towards the GC, BW, respectively. In comparison, the time-scale distribution
for NSs towards the BW based on WM05's model is shown as indicated.
}
\label{timescale}
\end{center}
\end{figure}
%%%%%%%%%%%%%%%%%%%%%%%%%%%%%%%%%%%%%%%%%

%
%%%%%%%%%%%%%%%%%%%%%%%%%%%%%%%%%%%%%%%%%
\begin{figure}
\begin{center}
  \includegraphics[width=3.5 in]{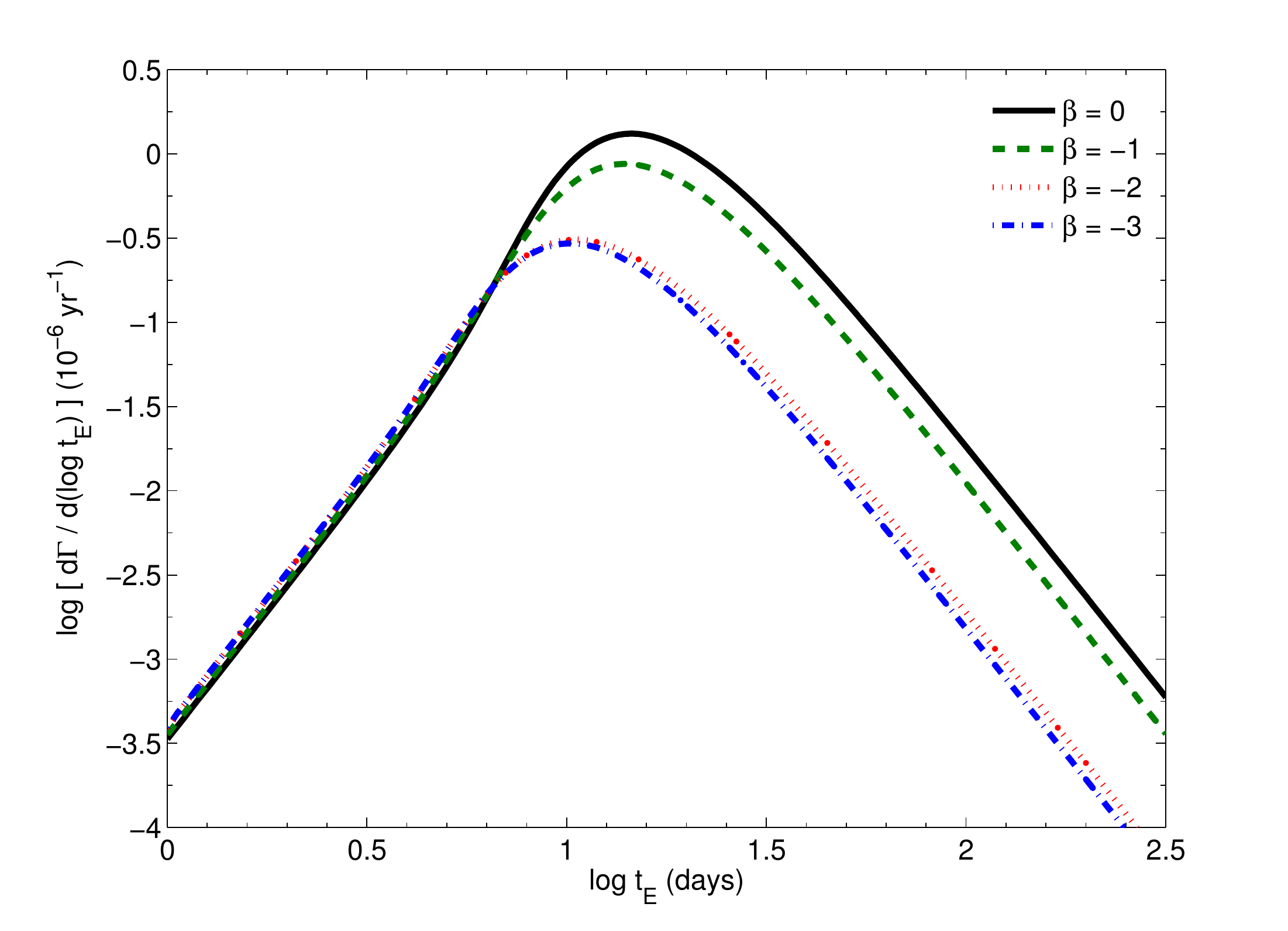}
\caption{Microlensing event rates caused by NSs as a function of time-scale towards
the GC. The solid, dashed, dotted and dot-dashed lines represent event rates for
$\beta=0,-1,-2,-3$, respectively.
}
\label{timescale_beta}
\end{center}
\end{figure}
%%%%%%%%%%%%%%%%%%%%%%%%%%%%%%%%%%%%%%%%%%
%
%
By restricting the Monte Carlo integrations in a parameter space corresponding to a certain time-scale, 
we can directly obtain the event rate caused by NSs as a function of time-scale. These are 
shown in Fig.~\ref{timescale} for $\beta = 0$. The time-scale distributions for the all-sky 
averaged event rate is shown as the solid line. The event rates towards the GC and BW 
are presented as dotted and dashed lines respectively. We note that the distributions 
have asymptotic power-law tails following the expected form given by~\citet{mao}. In all 
cases the event rate peaks at a time-scale of $\sim 12$\,days. The mean time-scales 
for the all-sky averaged, GC and BW event rates are $\sim 19.7$, $19.3$ and $19.4$\,days 
respectively. For comparison we also overlay the prediction from WM05 towards the BW which 
has a mean time-scale of $47.3$\,days. Our new results therefore imply that the majority 
of NS events will have shorter time-scales than predicted by WM05, which is mainly 
because of the relatively high velocity of NSs.
Using the exponential	model, we obtained a mean time-scale for the all-sky averaged event rate 
of $\sim 23.7$\,days.

In Fig.~\ref{timescale_beta}, we show the effects of changing $\beta$ on events 
towards the GC. 
The difference in the event rate between $\beta=-1$ and $-2$ is much larger than 
for other $\beta$ values because the background source density starts to decrease with
increasing distance according to the $D_{\rm{s}}^{2+2\beta}$ term in Eq.~\ref{rate}.
Below the mean time-scale the choice of $\beta$ has little effect on 
the results. $\beta$ does affect the longer time-scales with, as expected, smaller event 
rates as $\beta$ decreases. However, the mean time-scale is slightly dependent on $\beta$ 
with the mean time-scales being $19.3$, $17.9$, $13.5$ and $13.2$\,days for $\beta = 0$ to 
$-3$ respectively. This shift was not initially expected as decreasing $\beta$ reduces 
the averaged transverse velocity between the sources and lenses leading to an expected 
decrease in the number of short-duration events. In order to understand this result, 
we calculated the Einstein radii averaged over all lenses and the averaged transverse velocity 
between the lenses and sources. We found that, as $\beta$ decreased, the averaged Einstein 
radii decreases faster than the reduction in averaged transverse velocity. This implies 
that the resulting lensing events will occur by lenses with smaller Einstein radii and 
therefore the time-scale of these events will be shorter.

%the event rates decrease significantly and the time-scale
%distributions shift towards shorter durations.

\begin{figure}
\begin{center}
  \includegraphics[width=3.5 in]{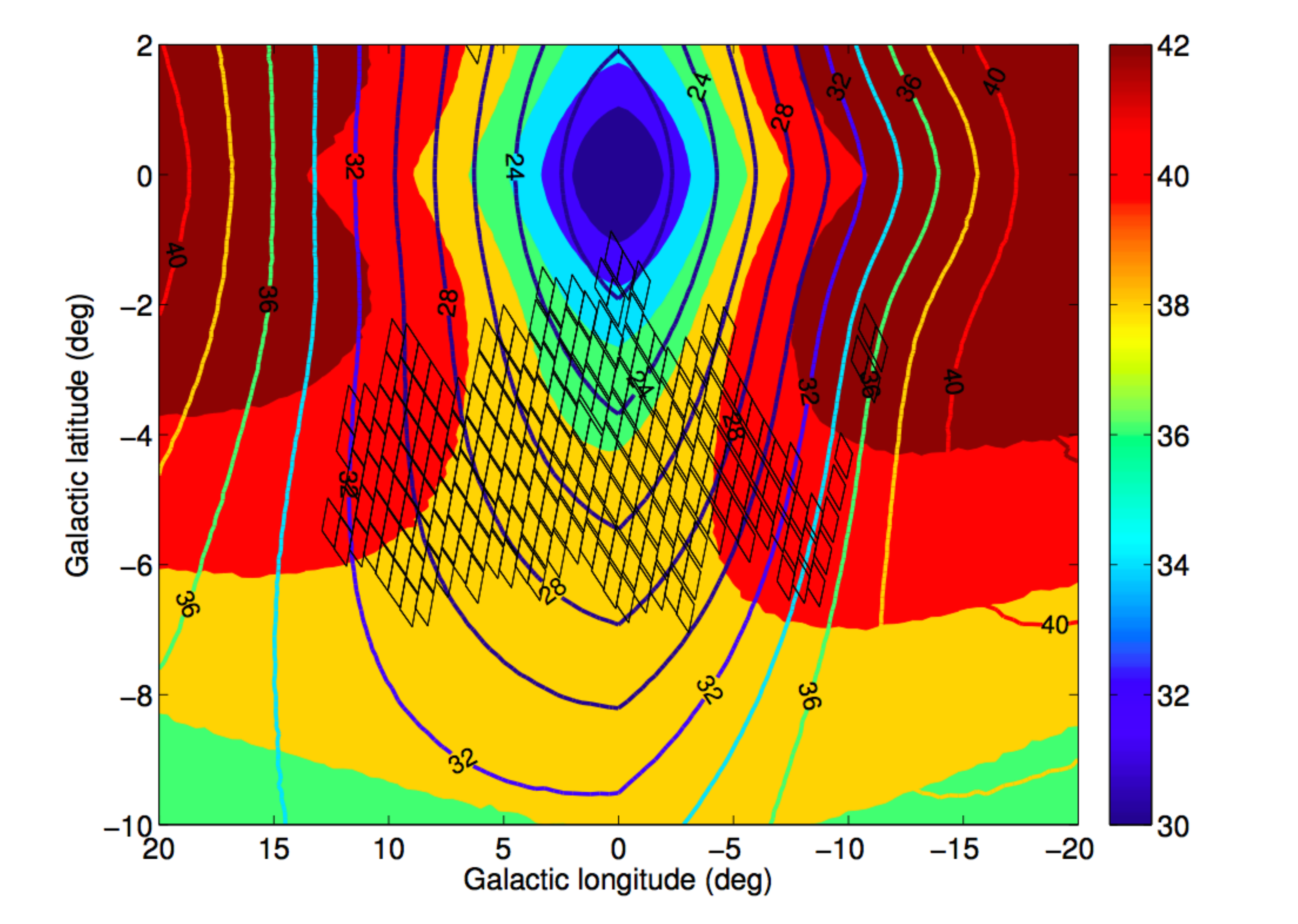}
\caption{Maps of the averaged time-scales of Galactic microlensing events. Contours with solid
lines represent results for $\beta=-1$. Contours filled with colors represent results for $\beta=-2$.
Contour levels represent averaged time-scales in unit of days. OGLE-III fields are shown with diamond symbols. }
\label{map_timescale}
\end{center}
\end{figure}

WM05 predicted the averaged time-scales of all (not just NS) Galactic events as a function 
of Galactic coordinates. The short duration events that we are predicting for NS lenses will 
lower the mean time-scale of these events. Our map is presented in Fig.~\ref{map_timescale} and 
can be compared with Fig.~3 in WM05. The solid contours represent $\beta = -1$ and the 
shaded regions as $\beta = -2$. The OGLE-III fields are shown with diamond symbols.  
%The WM05 map shows a flatter distribution that is elongated along the Galactic plane.	
Our averaged time-scales towards the BW are $23.3$, $23.9$ and $36.5$\,days for $\beta=0,-1,-2$, 
respectively. For $\beta=0,-1$, our results are similar to, but slightly smaller than, 
the corresponding results in WM05 and generally agrees with the actual measurement of 
$28.1\pm4.3$\,days~\citep{sumi} from the 
OGLE survey.

%%%%%%%%%%%%%%%%%%%%%%%%%%%%%%%%%%%%%%%%%
%
%%%%%%%%%%%%%%%%%%%%%%%%%%%%%%%%%%%%%%%%%%
%
%

\subsubsection{Fractional contribution to the Galactic event rate}

We have shown that NS events have relatively short time-scales. 
\citet{sumi11} showed that Jupiter-mass objects have even shorter time-scales,
while other stellar objects can have longer time-scales. In this subsection, we consider 
the fraction of Galactic events that are caused by NS lenses for different 
time-scales. 

\begin{figure}
\begin{center}
  \includegraphics[width=3.5 in]{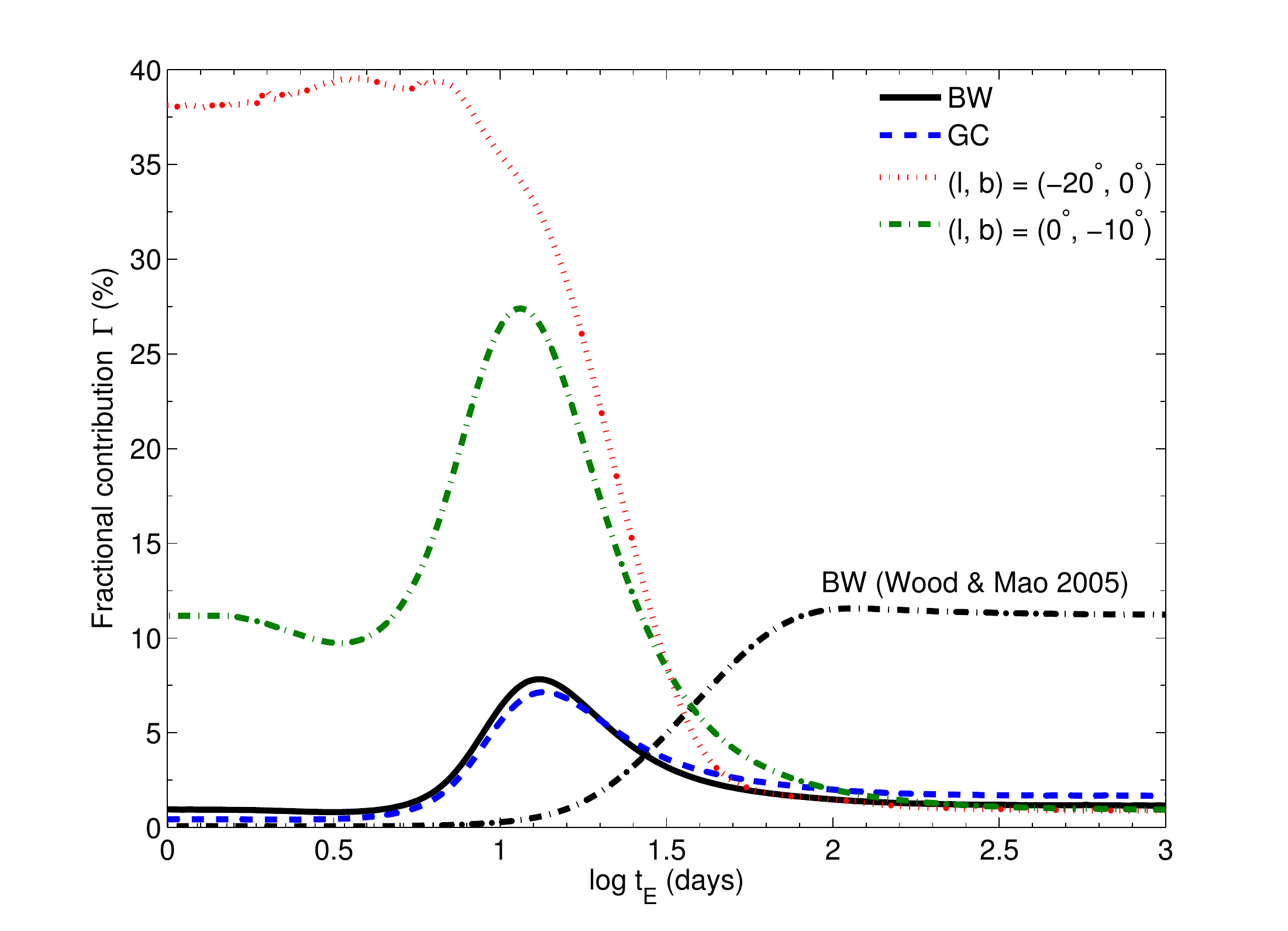}
\caption{Fractional contributions to the total Galactic event rate, as a function of
the event time-scale, from NSs. The solid, dashed, dotted lines and dot-dashed line show
the contribution from NSs towards the BW, GC and off-GC directions $(l,b)=(-20^{\circ},0^{\circ})$
and $(l,b)=(0^{\circ},-10^{\circ})$, respectively. In comparison, the fractional
contribution from NSs towards the BW based on WM05's model is shown as indicated.}
\label{ratio}
\end{center}
\end{figure}

In Fig.~\ref{ratio} we present the fraction of events that are caused by NSs as a 
function of time-scale. The solid line refers to events towards the BW and the dashed line 
towards the GC. We also show results for two off-GC directions, $(l,b)=(-20^{\circ},0^{\circ})$ 
and $(l,b)=(0^{\circ},-10^{\circ})$. For $(l,b)=(-20^{\circ},0^{\circ})$ (on the disk, but 
away from the bulge) NS lenses contribute up to $\sim$40\% of the events that last less 
than $\sim10$\,days, but we note that the total number of events in this direction is low. 
For the three other sky positions, the fractional contribution exhibits a clear peak around 
a time-scale of $\sim$15\,days and decreases at higher and lower time-scales. Our results 
do not agree with earlier predictions from WM05 (shown as a dot-dashed line in the Figure). 
Their work led to very small fractional contributions from NSs at
$\sim$15\,days and much larger fractional contributions on time-scales
of months to years.
On the contrary, in our calculations we have found that NSs mainly
contribute to short-duration Galactic events because of their
relatively high velocities (a factor not considered by WM05).
Away from the Galactic bulge region, as the rate of short-duration
Galactic events drops, the fractional contribution from NSs increases
rapidly at short time-scales.

\begin{figure}
\begin{center}
  \includegraphics[width=3.5 in]{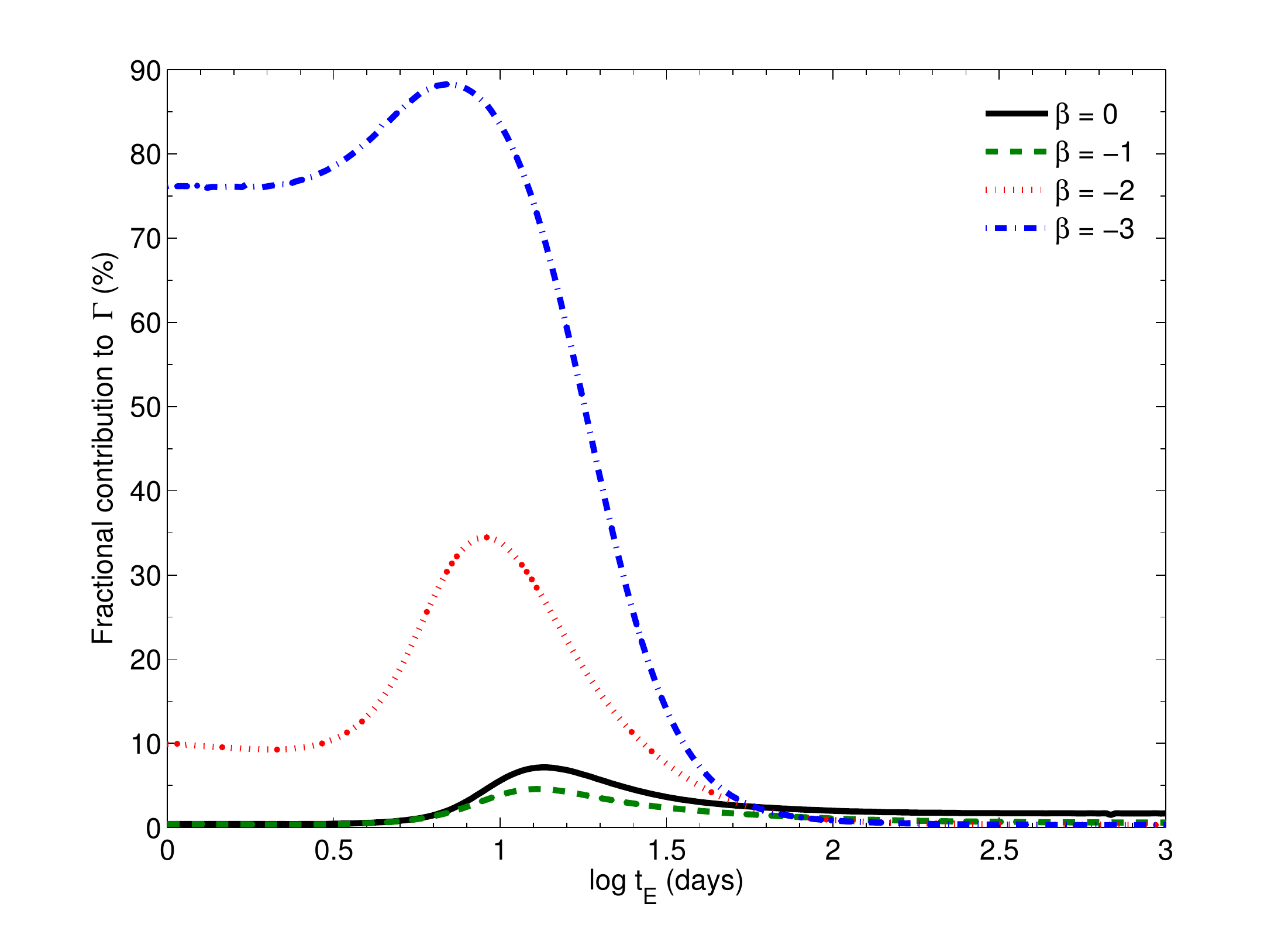}
\caption{Fractional contributions to the total Galactic event rate, as a function of
the event time-scale, from NSs, towards the GC. The solid, dashed, dotted and
dash-dotted lines represent the contributions for $\beta=0,-1,-2,-3$, respectively.}
\label{ratio_beta}
\end{center}
\end{figure}

In Fig.~\ref{ratio_beta} we show the fractional contribution towards the GC from NSs for 
different values of $\beta$. In all cases the fractional contribution peaks at $\sim 10$\,days 
and we cannot reproduce the WM05 results by modifying $\beta$. As $\beta$ decreases the 
fractional contribution on time-scales $\sim 10$\,days increases. For $\beta = -3$ almost 
all events at this time-scale would be caused by NS lenses. 

\begin{figure}
\begin{center}
  \includegraphics[width=3.5 in]{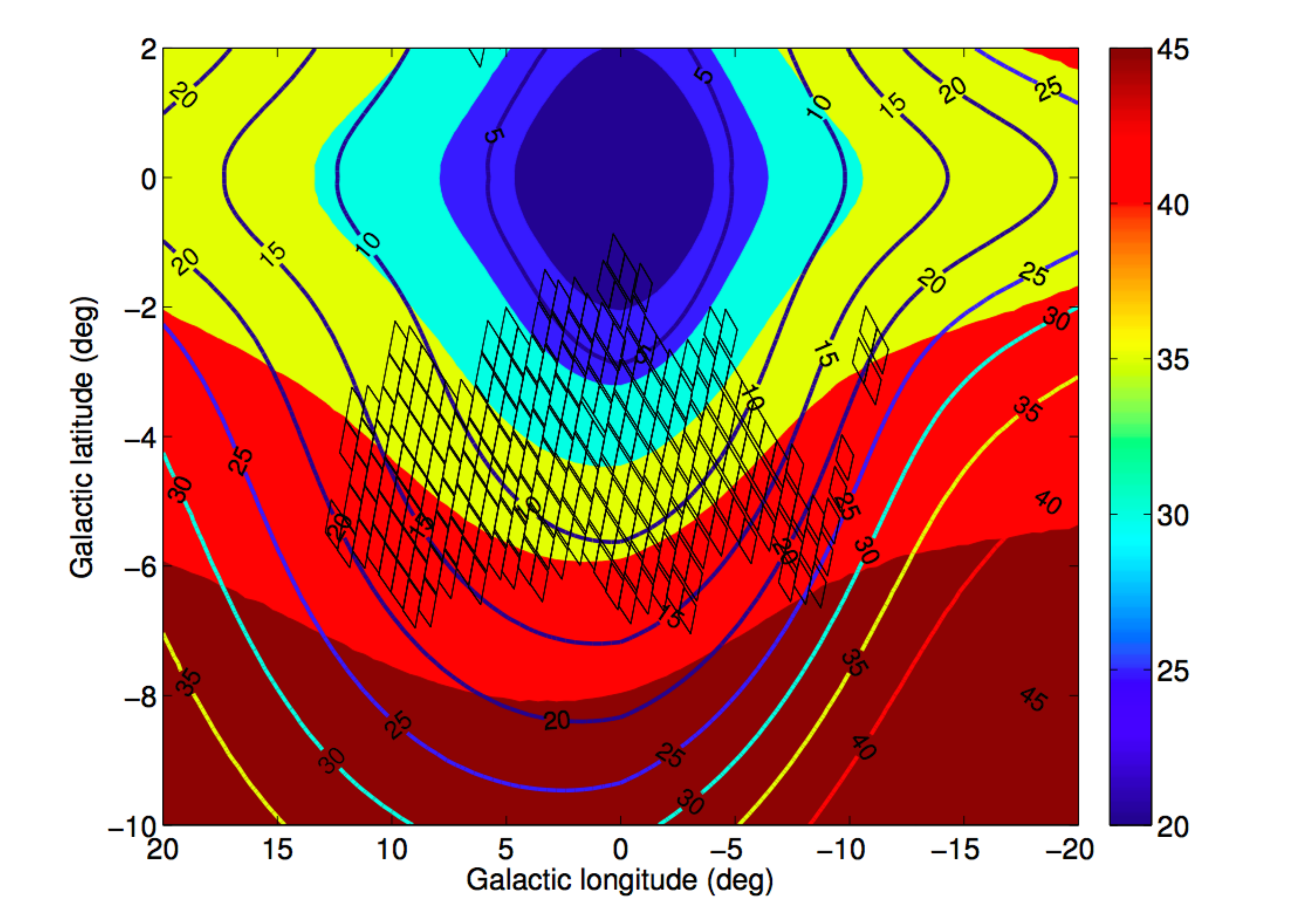}
\caption{Maps of the percentage contribution to the total Galactic microlensing event rates
from NSs. Contours with solid lines represent results for $\beta=-1$. Contours filled
with colors represent results for $\beta=-2$.  Contour levels represent the percentage contribution
in per cent. OGLE-III fields are shown with diamond symbols. }
\label{map_percentage}
\end{center}
\end{figure}

The fractional contribution from NS lenses changes as a function of sky position. Fig.~\ref{map_percentage} 
shows the fractional contributions from NSs in the central Galactic region. Contours indicated with 
solid lines are for $\beta = -1$. Contours shown as shaded regions are for $\beta = -2$. The OGLE-III 
fields are indicated using diamond symbols. The fractional contribution from NSs clearly varies 
with Galactic latitude with higher contributions from NSs occurring away from Galactic bulge. 

Averaged over all time-scales and over the entire sky we find that NSs contribute $\sim 5.4\%$ ($\beta = 0$) 
of the total number of Galactic events. In~Table~2, we present the fractional contribution to the total 
averaged event rate from different types of lensing object. The different objects considered are brown 
dwarfs (BDs), main-sequence stars (MSs), white dwarfs (WDs) and black holes (BHs). We also show  
contributions for different values of $\beta$. For $\beta = 0$, the NS contribution overall 
is small, however for other values of $\beta$ the contribution from NSs is much more significant.
By increasing and decreasing the $\sigma$ of the NS velocity distribution by 30\%, we 
got a fractional contribution from NSs of approximately 6.5\% and 3.7\% respectively ($\beta=0$).
Using the exponential	model, the fractional contribution from NSs is approximately 4.2\% ($\beta=0$).

%
%%%%%%%%%%%%%%%%%%%%%%%%%%%%%%%%%%%%%%%%%
%%%%%%%%%%%%%%%%%%%%%%%%%%%%%%%%%%%%%%%%%%
%
%
%
%%%%%%%%%%%%%%%%%%%%%%%%%%%%%%%%%%%%%%%%%
%%%%%%%%%%%%%%%%%%%%%%%%%%%%%%%%%%%%%%%%%%
%
%
%
%%%%%%%%%%%%%%%%%%%%%%%%%%%%%%%%%%%%%%%%%%%%%%%%%%%%%%%%%%%%%%%%%%%%%%%%%
\begin{table}
\begin{center}
\caption{Percentage contributions to the total event rate, from
different types of lens. Results for $\beta=-1,-2,-3$ are also presented.
In comparison, results based on WM05's model are also shown.}
\begin{tabular}{lccccc}
\hline
             &       \multicolumn{5}{c}{Types of lens}      \\
$\beta$      & BD   &    MS    &    WD     &  BH   & NS     \\
             &       \multicolumn{5}{c}{(per cent)}         \\
\hline
0            & 16.7 &    60.5  &    16.5   &  0.9  & 5.4    \\
-1           & 15.8 &    57.2  &    15.5   &  0.9  & 10.6   \\        
-2           & 10.6 &    38.2  &    10.4   &  0.5  & 40.3   \\         
-3           & 10.0 &    36.0  &    9.8    &  0.5  & 43.7   \\            
0 (WM05)     & 17.2 &    62.1  &    16.9   &  0.9  & 2.9    \\
\hline
\end{tabular}
\end{center}
\end{table}
%%%%%%%%%%%%%%%%%%%%%%%%%%%%%%%%%%%%%%%%%%%%%%%%%%%%%%%%%%%%%%%%%%%%%%%%%%
%
%
%%%%%%%%%%%%%%%%%%%%%%%%%%%%%%%%%%%%%%%%%
%%%%%%%%%%%%%%%%%%%%%%%%%%%%%%%%%%%%%%%%%%
%
%

\section{Mass measurement through astrometric microlensing}

We have investigated the properties of photometric microlensing events caused 
by NSs. However, the photometric microlensing event rate caused by potentially-observable 
radio pulsars is only $\sim2$ per decade for $10^{10}$ background sources. As astrometric 
microlensing has a larger cross-section than photometric microlensing, more astrometric 
microlensing events are expected to be detected. It is not currently feasible to carry out 
large scale astrometric microlensing surveys. A more promising way to search for events 
caused by observable radio pulsars is to use the known pulsar astrometric parameters to 
make an initial guess for when and where microlensing events will occur.  
This approach can also be applied to photometric microlensing, but the larger cross-section 
means that it is especially promising for astrometric detections.

The method we used to estimate the photometric microlensing event rate is also
valid for astrometric lensing, but the larger cross-section must be accounted for.  
The cross-section and the amplitude of an astrometric microlensing event can 
be characterized by the impact parameter, $u_0$, which is defined as the 
smallest angular separation between the lens and the source in units of the 
angular Einstein radius. The astrometric microlensing event rate will simply be 
our determination of the photometric event rate scaled by the increase in the 
cross-section. For instance, if the cross-section is $\sim 10$ times larger ($u_0=10$, 
discussed in more detail in Section 3.2) then, for 120,000 potentially-observable 
radio pulsars, the all-sky averaged event rate with $\beta=-1$ will then be $\sim1.5$ 
per year for $10^{10}$ background sources.

\subsection{Astrometric microlensing model}

BE02 presented methods to determine the precision that could be
achieved for stellar lens mass estimates obtained through astrometric
microlensing. They showed that there was a total of $11$ parameters to
completely parameterize such events.
We apply their methods to determine the precision achievable for observable radio pulsars.
%Their work focused on the \textit{Gaia} satellite and assumed that only one-dimensional 
%astrometry is available. We apply their methods to determine the precision achievable for 
%observable radio pulsars and assume a more general case in which two-dimensional astrometry is available.  

For radio pulsar lenses, the positions, proper motions and distances of lenses can, in 
principle, be measured or estimated through radio observations. Pulsar positions and proper 
motions can be determined either through radio pulsar timing or interferometric observations. 
For millisecond pulsars that have been observed over many years, the precision of positions 
and proper motions determined through pulsar timing could reach $\leq 1$\,mas and 
$\leq 1$\,$\rm{mas\ yr^{-1}}$, respectively~\citep[e.g.,][]{verbiest}.
For pulsars whose positions and proper motions cannot be obtained precisely through 
pulsar timing, interferometric observations can be used\footnote{We note that the dynamical 
solar system ephemeris used for pulsar timing is different from the International Celestial 
Reference Frame (ICRF). \citet{Madison13} computed a transformation between the ICRF 
and pulsar timing frames and quantitatively discussed how the transformation will improve in 
the coming years.}. In our model, we consider an ideal case in which we assume that the 
radio pulsar lens positions and proper motions are known parameters and we neglect their 
uncertainties. The implications of these assumptions are discussed at the end of Section~3.2.

Pulsar distances are much harder to measure than positions or proper motions.
For some pulsars precise distances can be obtained from measurements of parallax.
For most pulsars distances can only be estimated from dispersion measures. The conversion 
from dispersion measure to distance depends on a models for the distribution of 
free electrons in the Galaxy.
As such models are not very well constrained this results in distance uncertainties of
around 20\%~\citep{Taylor}.
Although models of the Galactic free electron density are continually being improved~\citep[e.g.,][]{cordes},
it is not clear whether distances will be measured sufficiently for most pulsars at 
the time that astrometric lensing events from pulsars will be detected.
Therefore, in our analysis we discuss two models. In the first model, which we 
label as Model $D_{\rm{psr}}^{\rm{known}}$, we assume that the pulsar distance is known. 
In the second model, Model $D_{\rm{psr}}^{\rm{unknown}}$ unknown we assume 
that it is not.

For Model $D_{\rm{psr}}^{\rm{known}}$, six independent parameters need to be computed from the data. 
These are the position ($\bm{\theta_{\rm{s,0}}}$), proper motion ($\bm{\mu_{\rm{s}}}$), 
and parallax ($\pi_{\rm{s}}$) of the background source and the mass ($M$) of 
the lens. We note that it is possible that the astrometric parameters for the background 
source is also precisely known. If true, then the only unknown parameter will be the mass.  
For this work, we assume that these parameters do need to be determined from the data. For  
Model $D_{\rm{psr}}^{\rm{unknown}}$, we have an additional parameter, the pulsar parallax ($\pi_{\rm{l}}$).  

As an example, we consider a specific pulsar, PSR~B1929$+$10. This pulsar has a large 
proper motion~\citep[103\,$\rm{mas\ yr^{-1}}$;][]{Chatterjee04} and lies in close projection 
to a number of potential background source stars, with seven objects within $5$\,arcsecs 
according to our catalog comparison between known radio pulsars\footnote{http://www.atnf.csiro.au/research/pulsar/psrcat}~\citep{Manchester05} 
and source tables from the UKIRT Infrared Deep Sky Survey~\citep[UKIDSS; ][]{Lawrence}.  
The distance to this pulsar is $0.36$\,kpc~\citep{Chatterjee04}. We assume that this pulsar 
is likely to pass one of the background sources and could be studied with an astrometric 
monitoring campaign that carries out observations of the background source with a uniform 
observing cadence over a period of four years (centered on the time of closest angular 
separation between the lens and the source). Following BE02, we assume that the background 
source is at a distance of 5\,kpc with a proper motion of 7\,$\rm{mas\ yr^{-1}}$.

We note that our selection of lens and source proper motions and distances will not 
limit the generality of our analysis, since the amplitude of an astrometric microlensing 
event is only characterized by $u_0$.  
However, the uncertainty on the mass measurements will depend on
various parameters that need to be assumed in our modeling. The first
is the pulsar mass. In our analysis, we use $M=1.4$\,$\rm{M_{\odot}}$
as a typical pulsar mass, but also trial different values. We also
require an estimate of the astrometric accuracy that can be achieved
for the background stellar object. We trial two possible values,
$\sigma=150$ and 450\,$\rm{\mu as}$. $\sigma=150$\,$\rm{\mu as}$
corresponds to a relatively bright star at around $V=16$ mag for the
\textit{Gaia} satellite\footnote{The end-of-mission parallax error for
such a star is around 20-40\,$\rm{\mu as}$. The single epoch
positional precision is significantly worse. We estimate how much
worse using information provided on
\url{http://www.cosmos.esa.int/web/gaia/science-performance} as
follows. The positional precision for a single epoch measurement is
more precise than the parallax precision by a factor of 0.743, but the
single-epoch precision is a factor of 4.3 times worse than that
available at the end-of-mission.  From this we estimate that the
single epoch positional precision is around three times worse.  We
also need to incorporate a factor of $\sqrt{2}$ to account for the
fact at each epoch \textit{Gaia} only measures the one-dimensional
position along the scanning direction. This gives an effective
two-dimensional astrometric precision of around 90 to 180\,$\rm{\mu as}$. 
Note that for faint stars ($V=20$\,mag) the end-of-mission parallax
precision drops to around 130 to 600\,$\rm{\mu as}$, meaning that the
effective two-dimensional astrometric accuracy will be around 0.5 to
3\,mas.}. Future telescopes such as the Wide-Field Infrared Survey
Telescope (WFIRST) have the potential to obtain astrometric precision
of a few tenths of a mas for even fainter
stars~\citep[e.g.,][]{Spergel}. Finally, the uncertainty also depends
on the number of epochs that are observed during the four
year monitoring program. In the results below we trial a range of
possible values.

\subsection{Mass determination results}

\begin{figure}
\begin{center}
  \includegraphics[width=3.5 in]{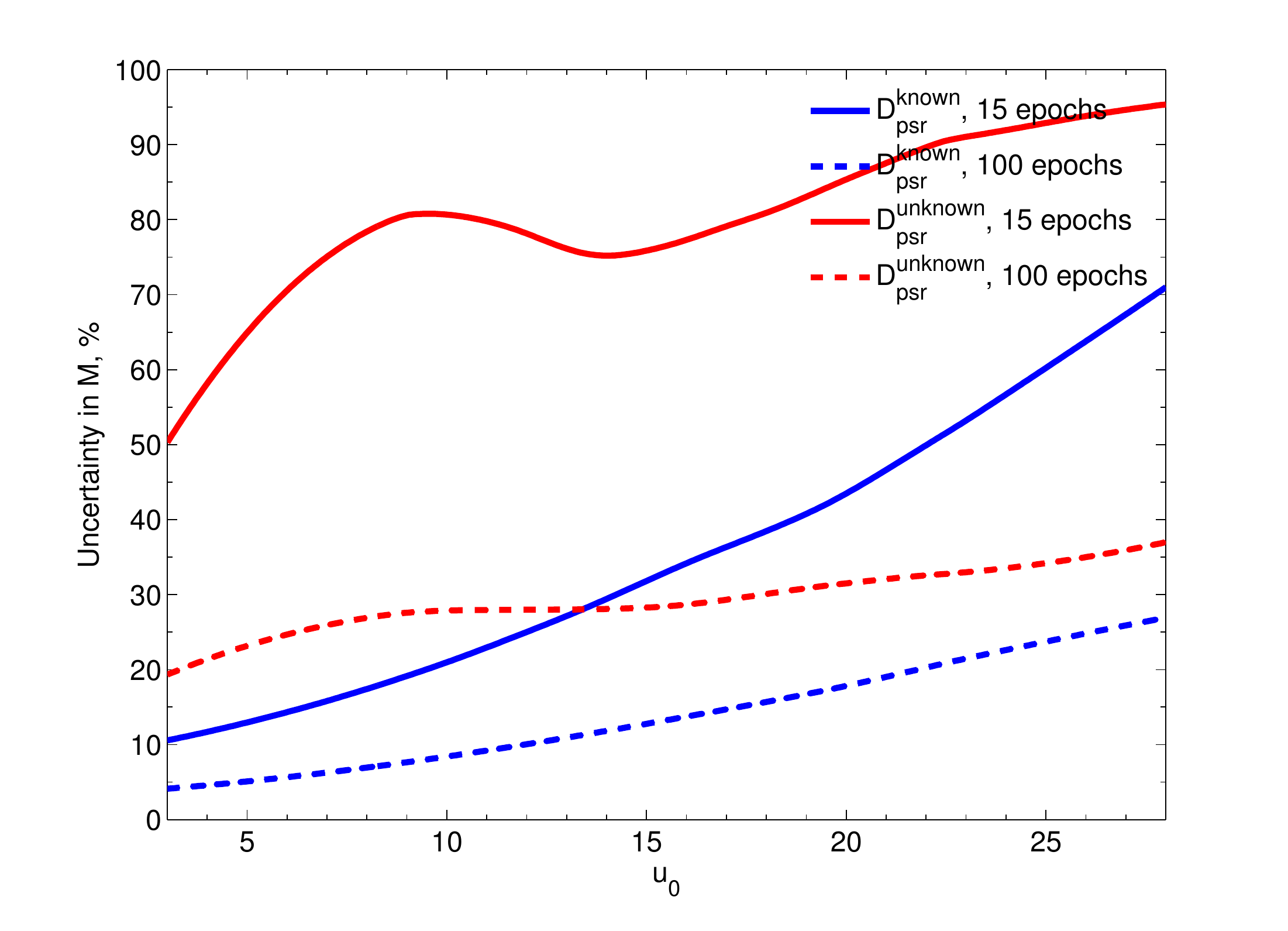}
\caption{The percentage error in estimation of $M$ as a function of impact parameter $u_0$.
We set $M=1.4$\,$M_{\odot}$ and $\sigma=150$\,$\rm{\mu as}$. The solid and dashed 
lines represent the errors with $15$ and $100$ observation epochs, respectively, and pulsar distances 
are supposed to be measured precisely (Model $D_{\rm{psr}}^{\rm{known}}$). While the dotted and dot-dashed lines represent 
the errors for which pulsar distances are set to computed from the data (Model $D_{\rm{psr}}^{\rm{unknown}}$).
}
\label{mass_u0}
\end{center}
\end{figure}

Fig.~\ref{mass_u0} shows the uncertainties in $M$ as a function of the dimensionless
impact parameter $u_0$ for the two models (in which the pulsar distance is known or not).  
We present results for our two models with different samplings (for 15 and 100 observation epochs). 
There is a large variation in the ability to determine the pulsar
mass. The uncertainties in $M$ increase with increasing $u_0$ and
decrease with the number of epochs. If the pulsar distance is known
then its mass could be determined at least two times better than if
the mass is not known. Note that for \textit{Gaia} the average number
of epochs is around 70 \citep{DeBruijne12}.
%
%
%%%%%%%%%%%%%%%%%%%%%%%%%%%%%%%%%%%%%%%%%
%%%%%%%%%%%%%%%%%%%%%%%%%%%%%%%%%%%%%%%%%%
%
%

\begin{figure}
\begin{center}
  \includegraphics[width=3.5 in]{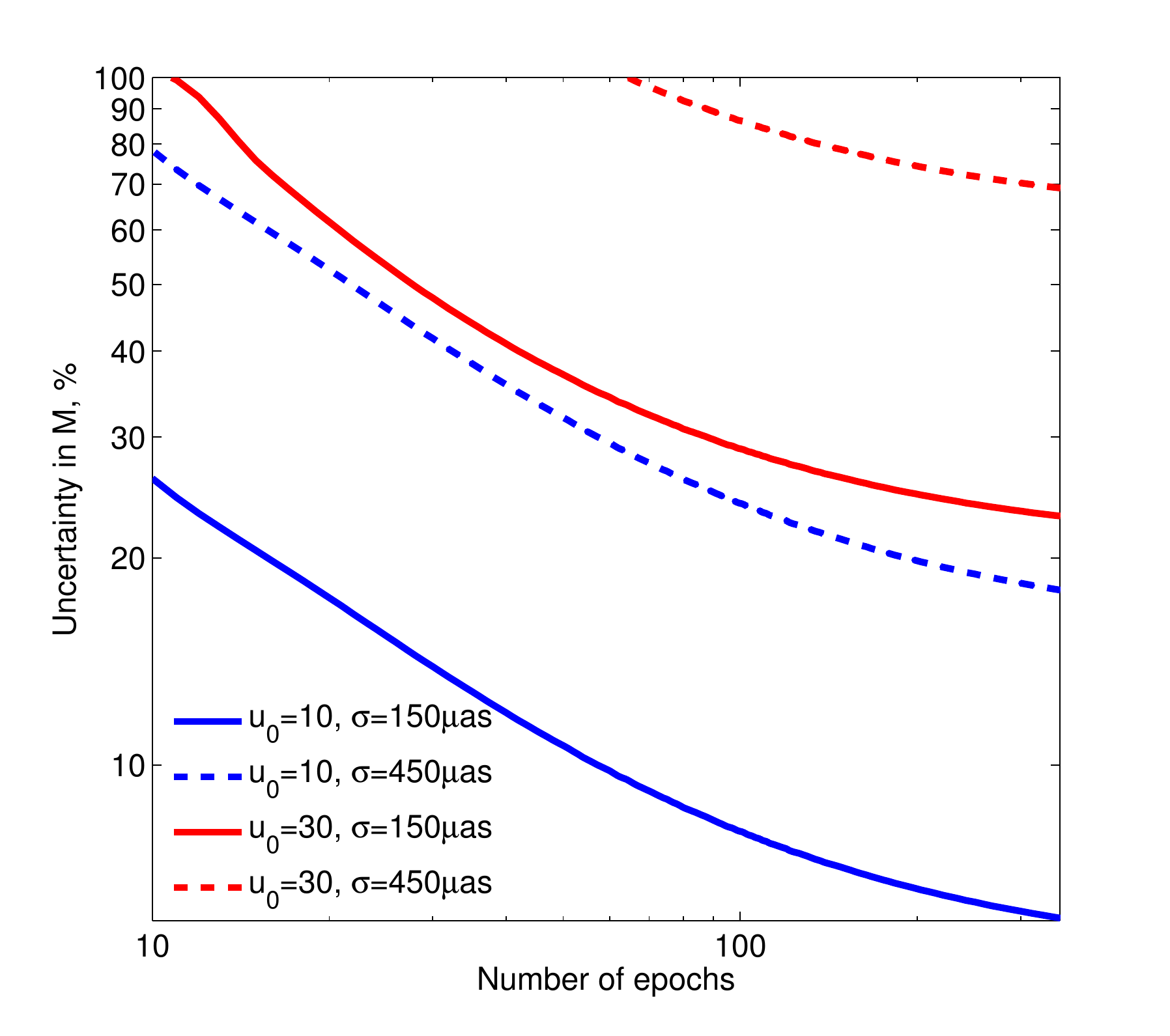}
\caption{The percentage error in estimation of $M$ as a function of the number of observation epochs,
for Model $D_{\rm{psr}}^{\rm{known}}$. Results for different $u_0$ and $\sigma$ are shown.
}
\label{mass_6}
\end{center}
\end{figure}

\begin{figure}
\begin{center}
  \includegraphics[width=3.5 in]{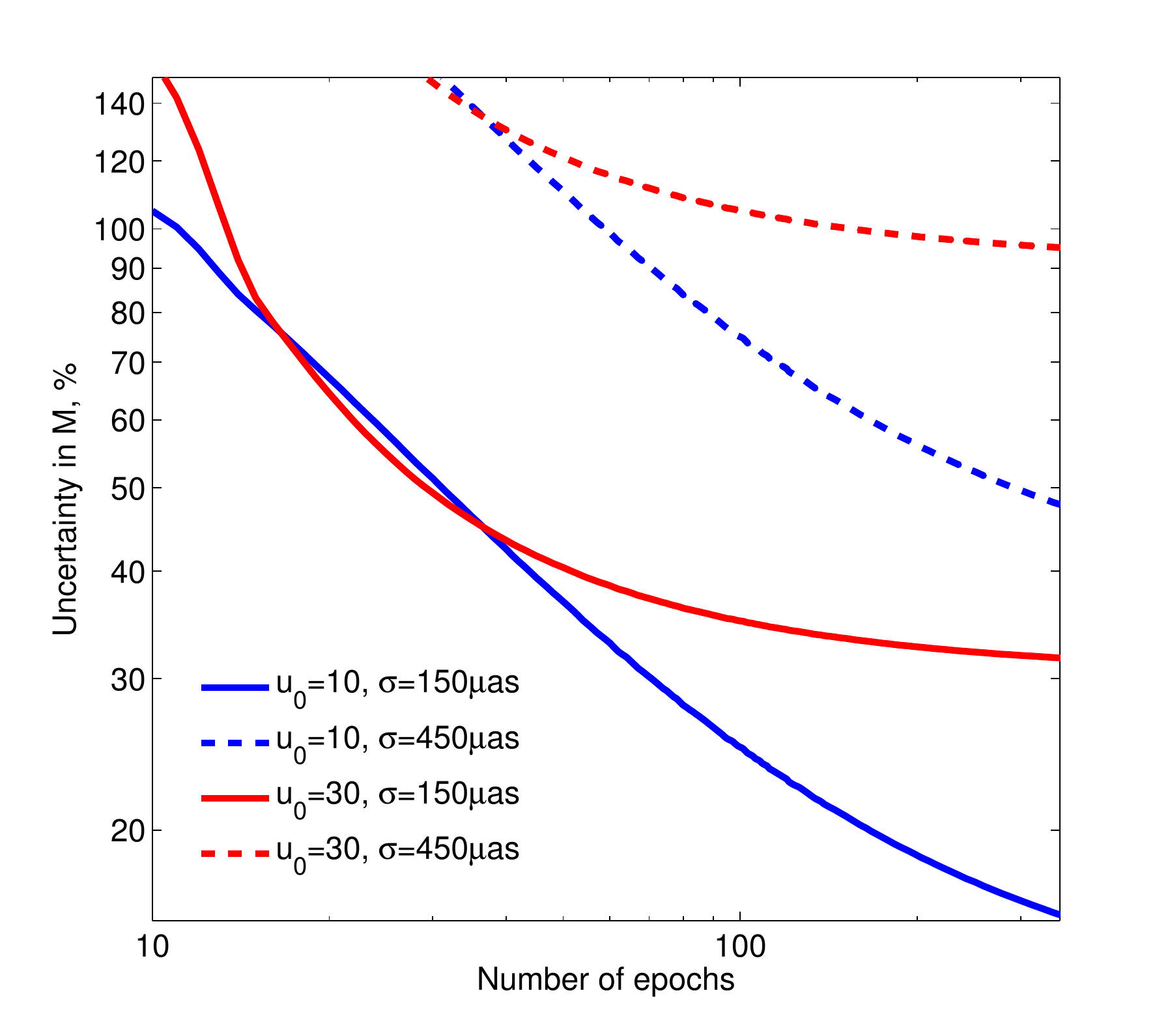}
\caption{The percentage error in estimation of $M$ as a function of the number of observation epochs,
for Model $D_{\rm{psr}}^{\rm{unknown}}$. Results for different $u_0$ and $\sigma$ are shown. }
\label{mass_7}
\end{center}
\end{figure}

Fig.~\ref{mass_6} and Fig.~\ref{mass_7} show the uncertainties in $M$ as a function
of the number of observation epochs. In Fig.~\ref{mass_6} we assume that the pulsar distance is known,  
while in Fig.~\ref{mass_7} we assume it is unknown. For both figures, we present results for 
$u_0=10$ and $u_0=30$, and for $\sigma=150$\,$\rm{\mu as}$ and $\sigma=450$\,$\rm{\mu as}$. 
%In Fig. \ref{mass_7}, we also show the uncertainties in $\pi_{\rm{l}}$ for $u_0=10$
%and $\sigma=150\ \rm{\mu as}$.
%
For more than $\sim 100$ epochs the uncertainty is approximately constant. More 
frequent sampling is therefore not necessary and will not improve the mass determination. 
As the astrometric precision improves from 450\,$\rm{\mu as}$ to 150\,$\rm{\mu as}$, 
the uncertainty in $M$ decreases by more than a factor of two. Clearly, it is therefore 
essential that the background source is monitored with high precision astrometry. 
%
%%%%%%%%%%%%%%%%%%%%%%%%%%%%%%%%%%%%%%%%%
%%%%%%%%%%%%%%%%%%%%%%%%%%%%%%%%%%%%%%%%%%
%
%
%%%%%%%%%%%%%%%%%%%%%%%%%%%%%%%%%%%%%%%%%
%%%%%%%%%%%%%%%%%%%%%%%%%%%%%%%%%%%%%%%%%%
%

\begin{figure}
\begin{center}
  \includegraphics[width=3.5 in]{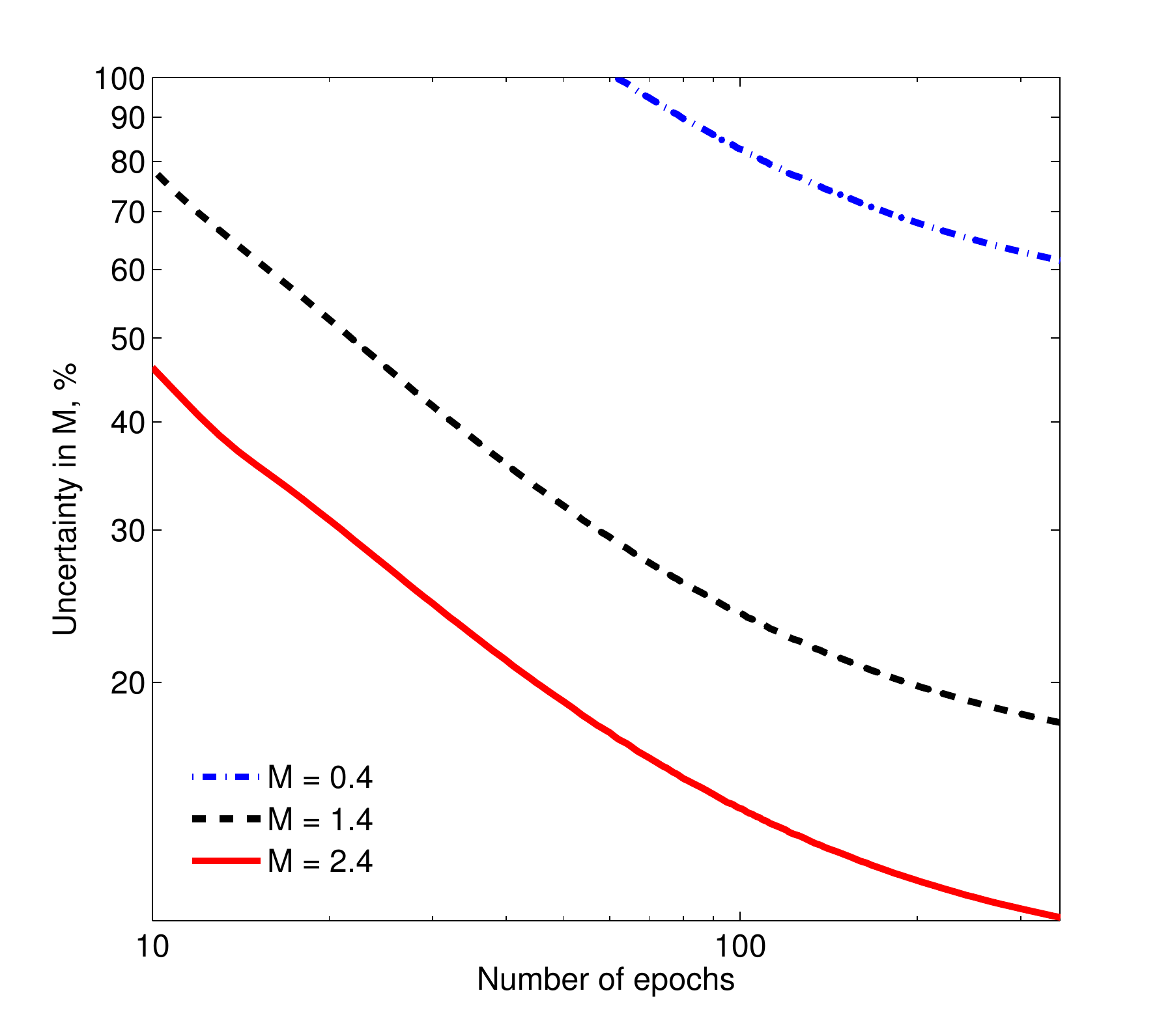}
\caption{The percentage error in estimation of $M$ as a function of the number of observation epochs,
for Model $D_{\rm{psr}}^{\rm{known}}$. Results for the pulsar mass of 0.4, 1.4 and 2.4\,$\rm{M_{\odot}}$ are shown.
}
\label{masses}
\end{center}
\end{figure}

In Fig.~\ref{masses}, we set different pulsar masses and show the uncertainties in 
$M$ as a function of the number of epochs for Model $D_{\rm{psr}}^{\rm{known}}$. In addition to a typical pulsar 
mass of 1.4\,$\rm{M_{\odot}}$, we also show results for an extremely low-mass case, 0.4\,$\rm{M_{\odot}}$, 
and a relatively high-mass case, 2.4\,$\rm{M_{\odot}}$. We set $u_0=10$ and 
$\sigma=450$\,$\rm{\mu as}$. 
As the pulsar mass increases the uncertainties become smaller. For 
the high-mass case, $M=2.4$\,$\rm{M_{\odot}}$, the uncertainty is six times smaller 
than that of the low-mass case, $M=0.4$\,$\rm{M_{\odot}}$.
%
%The reason is that for high mass lens the astrometric microlensing effects are 
%larger and then we can better measure the lens mass.
%
%%%%%%%%%%%%%%%%%%%%%%%%%%%%%%%%%%%%%%%%%
%%%%%%%%%%%%%%%%%%%%%%%%%%%%%%%%%%%%%%%%%%
%
%

In the analysis above, we have presented results for an ideal case in which we treat pulsar 
positions and proper motions as precisely determined values and we neglect their errors. If the 
pulsar astrometric parameters are poorly known then those parameters would need to be included in 
the model when determining the lensing event properties. This would significantly degrade the 
mass determination precision. Here, we investigate an intermediate case in which the astrometric 
parameters have been measured, but do have small uncertainties. We therefore model them as 
additional astrometric uncertainties of background sources and then apply the same method 
described above to estimate the uncertainty in pulsar masses.
We assume that the uncertainties on the pulsar position and proper motion measurements are 
0.01\,$\rm{mas}$ and 0.1\,$\rm{mas\ yr^{-1}}$ respectively. This is similar to that obtainable 
with current Very Long Baseline Interferometry (VLBI) results~\citep[e.g.,][]{deller}.  
For Model $D_{\rm{psr}}^{\rm{known}}$, assuming $u_0=10$ and $50$ epochs, the uncertainty 
on the mass measurement increases slightly from $10\%$ to $12\%$ compared with our earlier analysis.
For larger pulsar position and proper motion errors of 0.1\,$\rm{mas}$ and 1\,$\rm{mas\ yr^{-1}}$,
the uncertainty of mass measurement increases to $22\%$.
Therefore, we conclude that the errors of pulsar position and proper motion can
introduce additional uncertainties to pulsar mass measurements, 
but the effect is relatively small and can be significantly reduced by precise 
pulsar position and proper motion measurements already achievable with current VLBI observations.

\section{Conclusions}

A number of current and future surveys are carrying out photometric monitoring campaigns of 
the Galactic bulge and disc, such as OGLE-IV, Vista Variable in the Via Lactea~\citep{Minniti} and 
WFIRST~\citep{Spergel}. In the near future the LSST~\citep{ivez} 
will monitor $\sim 10,000$ square degrees of sky using pairs of $15$-second exposures twice per night 
every three nights on average. This will lead to a typical $5\sigma$ depth for point sources of $r\sim24.5$ (AB). 
Currently it is unclear to what extent LSST will survey the Galactic disc and bulge. Even without large portions 
of the bulge and disc, LSST will monitor $10$ billion Milky Way stars and, if a strategy is chosen to 
include the disc and bulge, then this number will increase significantly. A specific $3$-year HST program 
($192$ orbits) to detect microlensing events caused by non-luminous isolated black holes and NSs in 
the direction of the Galactic bulge has also been proposed~\citep{sahu}. 

Our modeling has shown that:
\begin{itemize}
\item The time-scale of microlensing events caused by NSs is much shorter ($\sim 20$\,d) than previously thought.
\item Events near the Galactic centre that have a time-scale of $\sim 15$\,d will contribute to approximately 7\% of the total number of events. Away from the bulge region the fractional contribution, for events with time-scales of $< 20$\,d will increase to around 40\%.
\item For an astrometric microlensing event caused by a known radio pulsar, pulsar masses could be determined with a precision of $\sim 10$\% if pulsar distances are known. With a sufficient number of observations of the event, the mass could be measured to $\sim 25$\% if the distance is not known.
\end{itemize}

For ongoing and upcoming surveys, we recommend:
\begin{itemize}
\item To maximize NS events in the microlensing dataset, surveys of the Galactic disc and bulge should be carried out.
\item NS events should be identified by searching for short duration ($\sim 20$\,d) events (instead of long duration events that had been previously considered) and radio telescopes should observe the possible candidates in order to identify radio pulsars.
\item As new pulsars are discovered, their astrometric parameters should be determined precisely and compared with catalogs of stellar positions. If a lensing event is likely then an astrometric lensing monitoring campaign should be carried out around the predicted time of the event.
\item Although these targeted campaigns are clearly the best approach to
detect astrometric microlensing events from potentially-observable
radio pulsars, the prospects for ESA's \textit{Gaia}
mission~\citep{DeBruijne12} are not negligible. In total \textit{Gaia}
will observe $\sim10^9$ stars and so, according to our estimates, this
equates to around one such event over the lifetime of the mission
(five years).
\end{itemize}

In the more distant future, large radio telescopes will have discovered most of the potentially-observable 
radio pulsars in the Milky Way. 
%Our modelling has shown that the all-sky averaged event rate for these radio 
%pulsars will be $\sim 7.2 \times 10^{-11}$\,yr$^{-1}$ for each background source. In the SKA era 
It is therefore likely that the mass of isolated NSs will be measurable through 
microlensing events leading to improved knowledge of the pulsar population. Extremely high-mass and 
low-mass pulsars can then be studied in detail to provide new insight into the NS interior.

\acknowledgments

We would like to acknowledge useful discussions with Dr. Jian Wang and
members of pulsar group at PKU.
SD is supported by China Scholarship Council (CSC).
M.C.S. acknowledges financial support from the CAS One Hundred Talent
Fund and NSFC Grant 11173002 and 11333003. This work is partially supported by the
Gaia Research for European Astronomy Training (GREAT-ITN) Marie Curie
network, funded through the European Union Seventh Framework Programme
(FP7/2007-2013) under grant agreement no 264895.
GH is a recipient of a Future Fellowship from the Australian Research Council.
This work is also supported by the National Natural Science Foundation of
China (11225314), the National Basic Research Program of
China (2012CB821801), the Strategic Priority Research Program "The Emergence of 
Cosmological Structures" of the Chinese Academy of Sciences, Grant No. XDB09000000.
This research has made use of NASA's Astrophysics Data System.

\bibliography{ms}
\end{document}